\documentclass[11pt]{article}
\usepackage{amssymb}

\newcommand{\half}{\mbox{$\frac{1}{2}$}}
\newcommand{\fslash}{\!\!\not\!}
\newcommand{\dslash}{\!\!\not\!\partial}
\newcommand{\tr}{\mbox{tr}}

\begin{document}

\begin{table}
\begin{flushright}
IK--TUW 9909401
\end{flushright}
\end{table}

\title{Bosonization in four dimensions: The smooth
way\thanks{Supported by Fonds zur F\"{o}rderung der wissenschaftlichen
Forschung, P11387--PHY}}

\author{Jan B.\ Thomassen\thanks{E-mail: {\tt
thomasse@kph.tuwien.ac.at}} \\
{\em Institut f\"{u}r Kernphysik, Technische Universit\"{a}t Wien} \\
{\em A--1040 Vienna, Austria}}

\date{November 16, 1999}

\maketitle

\begin{abstract}

I investigate bosonization in four dimensions, using the smooth
bosonization scheme. I argue that generalized chiral ``phases'' of the
fermion field corresponding to chiral phase rotations and ``chiral
Poincar\'{e} transformations'' are the appropriate degrees of freedom
for bosonization. Smooth bosonization is then applied to an Abelian
fermion coupled to an external vector. The result is an exact
rewriting of the theory, including the fermion, the bosonic fields,
and ghosts. Exact bosonization is therefore not achieved since the
fermion and the ghosts are not completely eliminated. The action for
the bosons is given by the Jacobian of a change of variables in the
path integral, and I calculate parts of this. The action describes a
nonlinear field theory, and thus static, topologically stable solitons
may exist in the bosonic sector of the theory, which become the
fermions of the original theory after quantization.

\vspace{\baselineskip}
\noindent
PACS numbers: 11.15.Tk; 11.30.Rd; 11.10.Lm \\
{\em Keywords}: Bosonization; Chiral anomalies; Nonlinear field
theories

\end{abstract}

\section{Introduction}

Bosonization is an operation which maps the description of a physical
system in terms of fermionic fields into a description in terms of
bosonic fields. Such operations have been known in relativistic field
theory since the seminal paper by Coleman \cite{coleman}, and in
condensed matter physics even for a longer time \cite{mattis}. A more
precise statement of bosonization is that there is a correspondence
between a certain set of Green's functions in the fermionic theory
with another set in the bosonic theory.  Reasons for wanting to
bosonize a fermion are for example that the description of the system
might be easier to handle mathematically, that it brings insight into
the physics of the system, or that it is otherwise useful.

However, bosonization is understood only in two dimensions (2D); it is
not yet completely understood how to generalize it to four. Some
papers dealing with this problem in the path integral formalism are
refs.\ \cite{diakonov,damgaard1,damgaard2,burgess} and references
therein. An investigation in the operator formalism is\footnote{I
thank J.\ Paaske and the referee for calling my attention to this
reference.} \cite{luther}; see also the references in
\cite{burgess}. The results from these papers do not coincide. In the
case of the path integral formalism this is a reflection of the fact
that a number can be represented by an integral in many different
ways. Therefore, bosonization is not unique, and further requirements
must be formulated in order to proceed.

What I will do in this paper is to investigate bosonization in the
path integral in 4D from another point of view, different from those
in the references above, but which perhaps follows more closely the
ideas in 2D from Coleman's paper \cite{coleman}.  Let us recall that
in 2D the bosonic equivalent of a massive Dirac fermion is the
sine--Gordon model (with coupling constant $\beta=\sqrt{4\pi}$), a
nonlinear field theory which has static, topologically stable
solitons. One of the main messages of ref.\ \cite{coleman} is that
these solitons have properties which allow us to identify them with
fermions in the quantum theory, and indeed Mandelstam
\cite{mandelstam} showed in the operator formalism that fermionic
operators satisfying the Dirac equation (for the Thirring model) can
be constructed from the bosonic operators satisfying the sine--Gordon
equation.

Thus, we also want something like this in 4D: A nonlinear bosonized
theory with static, topologically stable solitons that can be
identified with the fermions in the original theory. For this to be at
all possible in three space dimensions, it is necessary for
topological reasons to have bosonic fields with at least three degrees
of freedom. Furthermore, it is necessary to have a mechanism which
prevents the solitons from collapsing according to Derrick's theorem
\cite{derrick}, like terms with four derivatives in the Lagrangian,
which is the situation in the Skyrme model \cite{skyrme}. We also want
a Skyrme-like picture to be possible for an Abelian Dirac
fermion. This is different from the Skyrme model where there is a
triplet of pseudoscalar fields which transforms under an internal
symmetry group (the pion), and where the fermion (the nucleon) acquire
a corresponding internal quantum number. I am, however, not aware of
any reasons why the internal symmetry could not be spin, leading to a
Skyrme soliton with a $U(1)$ quantum number.

Note that if we express the equivalence between the fermionic and
bosonic descriptions in terms of Green's functions, then only Green's
functions of bilinear operators of fermions could be considered, like
in 2D. This may seem to imply that some information is lost by
bosonization -- namely the information in Green's functions with
external spinor lines. However, this is not necessarily the case. In
2D this information is encoded in the solitons of the bosonic theory,
as is verified from Mandelstam's construction of the Dirac field
operators \cite{mandelstam}. This is one way to understand our
requirement of 4D bosonization that Skyrme-like fermionic solitons
should appear in the theory: There is then a chance that spinor
information is encoded in the bosonic theory.

Another thing we recall from 2D is that in the path integral
investigations of bosonization the bosonic field equivalent to the
Dirac fermion is essentially the chiral phase of the
fermion. Moreover, the Lagrangian for the boson is in one way or
another connected with the Jacobian of a chiral phase rotation of the
fermion (see for instance ref.\ \cite{abdalla}), that is, if we write
the Jacobian as $J=\exp(i\int d^2x{\cal L}_J)$, then the Lagrangian is
connected with ${\cal L}_J$. The exact connection depends on the
particular bosonization scheme, but ${\cal L}_J$ always contains at
least the kinetic terms of the bosonic fields. Thus, in 4D we should
be looking for something similar to the chiral phase of the fermion,
which gives rise to a Jacobian when the fermion is transformed, and
such that this Jacobian can be connected with the Lagrangian for the
bosonic field.

To summarize our requirements, we want:

(i) ``generalized chiral phases'' of the fermion, which should be
fields with at least three internal degrees of freedom, like a vector
or a tensor,

(ii) that the ``generalized chiral transformations'' for the fermion
give rise to a Jacobian when the variables are changed in the path
integral, and when a suitable regularization scheme is applied, and

(iii) the Jacobian $J=\exp(i\int d^4x{\cal L}_J)$ must be such that
the Lagrangian quantity ${\cal L}_J$ contains, at least, the kinetic
terms of the chiral fields, and also other terms with four derivatives
that can stabilize solitons in the bosonic theory.

I emphasize that these are not the only requirements one can make for
4D bosonization, but I think they are a reasonable starting point if
we want to generalize from 2D the idea that solitons are fermions in
the quantum theory.

As we shall see, we are able to meet these requirements when the
generalized chiral phases are degrees of freedom (DOFs) connected with
what may be called ``chiral Poincar\'{e} transformations''
\cite{thomassen}. These are (global) transformations that act on the
Dirac fermion in the same way as ordinary Lorentz transformations and
translations, except that a $\gamma_5$ is multiplied onto the
generators $J_{\mu\nu}=\half\sigma_{\mu\nu}+(x_\mu i\partial_\nu-x_\nu
i\partial_\mu)\equiv S_{\mu\nu}+L_{\mu\nu}$ and
$P_\mu=i\partial_\mu$. The reason I consider exactly {\em these} DOFs
is the following: The particle states in a relativistic quantum field
theory are labelled by internal quantum numbers, spin, and mass,
according to which representation of the internal symmetry group and
Poincar\'{e} group they belong. Since these properties are connected
with ``phase rotations'', Lorentz transformations, and translations,
and since the chiral phase is likely to take part in bosonization,
then it is also likely that chiral Lorentz DOFs and chiral translation
DOFs take part as well.  Such transformations are not in general
symmetries of the Lagrangian, but naively they are symmetries of the
fermionic measure in the path integral. However, in ref.\
\cite{thomassen} it was shown that they give rise to a Jacobian when a
change of fermionic variables is made in the {\em regularized} path
integral. Moreover, the fields connected with these DOFs are an axial
vector and a tensor, which may have nontrivial topology. Thus, we may
regard the quantity ${\cal L}_J$ for these chiral transformations as a
candidate for the Lagrangian of a bosonized theory.

It is then necessary to show that this bosonic theory really is
equivalent to the original fermionic one, or that it is in some sense
a partial bosonization.  In 2D an appealing method to do bosonization
in the path integral is the ``smooth bosonization scheme'' of
Damgaard, Nielsen and Sollacher \cite{damgaard3,damgaard4}. This is
the scheme I will adopt here for 4D -- hence the title of this
paper. (This scheme has already been used to obtain partial
bosonization in 4D for some special cases \cite{damgaard1,damgaard2}.)
Briefly stated, the method is to perform a change of variables in the
path integral, using instead a fermion that is locally rotated with a
chiral phase $\theta(x)$. This gives a Jacobian $J[\theta]=\exp(i\int
d^2x{\cal L}_J[\theta])$, assuming an appropriate regularization is
used. The field $\theta$ enters into the expression for the path
integral, and the theory is enlarged by promoting it to a dynamical
field, meaning that we path integrate over it. The theory now has a
new local symmetry. This symmetry can be gauge-fixed, and by carefully
choosing the gauge-fixing condition, we can get either the original
fermionic theory or a new bosonic theory -- the bosonized theory. The
Lagrangian for this theory is ${\cal L}_J[\theta]$. I emphasize that
this bosonization scheme is based on {\em exact} manipulations of the
path integral, and uses only familiar concepts from field theory -- in
particular gauge-fixing and chiral anomalies.

In this paper, I will investigate the bosonization of a massive
Abelian Dirac fermion, coupled to an external vector field, along
these lines.  I will show how the new ``generalized chiral phases'' of
the fermion, corresponding to chiral Poincar\'{e} transformations, can
be introduced into the path integral. Then I will discuss how the new
local symmetry of the path integral can be gauge-fixed, and I will use
the gauge-fixing constraints for the chiral phase and chiral
Poincar\'{e} symmetry, along with the conservation of the current,
energy-momentum and angular momentum tensors, to try to eliminate the
fermion from the theory. The resulting theory includes a Lagrangian
${\cal L}_J$ for a pseudoscalar $\theta$, a tensor $\phi_{\mu\nu}$ and
an axial vector $b_\mu$. These fields are the parameters of the local
chiral transformations. Unfortunately, the ghosts do not decouple in
the chosen gauge, and, furthermore, a ``small'' residual part of the
fermion is not eliminated. Nevertheless, the resulting theory may
still be useful. Perhaps it can serve as a starting point for
obtaining a low energy effective description of the system.

The organization of the paper is the following. In sec.\ 2, I briefly
review the smooth bosonization scheme \cite{damgaard3}. The scheme was
first applied to a massless fermion; the massive case was considered
in ref.\ \cite{damgaard4}. However, the treatment of the massive case
in ref.\ \cite{damgaard4} was not as straightforward as for the
massless case. In particular, the simple result that the bosonic
Lagrangian equals ${\cal L}_J$ was not found. I therefore reconsider
this massive case, and find that one may indeed achieve that the
bosonic Lagrangian is ${\cal L}_J$. What makes this work can be
understood as the effective vanishing of the axial current (and
therefore also the current), which is a consequence of the combined
effects of gauge-fixing and current conservation.
 
In sec.\ 3, I begin my investigations of a massive fermion coupled to
a vector in 4D. First I introduce the chiral phase $\theta$ into the
theory, inspired by the 2D case. The appropriate gauge-fixing for this
DOF, together with energy-momentum and angular momentum conservation,
is used to try to make the axial current effectively vanish. Then I
introduce the chiral Lorentz and chiral translation ``phases'',
$\phi_{\mu\nu}$ and $b_\mu$, into the theory in an analogous way and
use gauge-fixing of these DOFs, together with current conservation, to
try to get the effective vanishing of the current. The idea is that
the vanishing of both the current and the axial current should be
enough to eliminate the fermion from the theory. The exact theory
which results from this is given by a path integral with the original
fermion, the bosonic fields (including Lagrange multipliers for the
gauge-fixing delta functions), and ghosts.

In sec.\ 4, I discuss the calculation of the Jacobian of the chiral
transformation, which would describe the bosonic theory in the best of
all worlds where bosonization is exact. The complete Lagrangian ${\cal
L}_J$ is probably too hard to calculate exactly. However, the part
containing only $\theta$ {\em can} be calculated exactly. I also
calculate the Gaussian part (i.e.\ up to two orders in the field) of
the spin part of the action for $\phi_{\mu\nu}$. Terms involving
$L_{\mu\nu}$ and the Lagrangian for $b_\mu$ (which involves $P_\mu$)
are not calculated. The terms that are left out are thereby small in
the weak field, low energy limit. The lowest order coupling terms
between $\theta$ and $\phi_{\mu\nu}$ are also calculated.

In sec.\ 5, I summarize and discuss the problems which prevented us
from getting exact bosonization in sec.\ 3. I sketch how to generalize
the results to the non-Abelian case and discuss some aspects of the
physics of the bosonized theory.

\section{Smooth bosonization in 2D: The massive case revisited}

The smooth bosonization scheme has been used in 2D to bosonize a Dirac
fermion coupled to a vector and axial vector \cite{damgaard3}, and a
Dirac fermion coupled to scalar and pseudoscalar mass terms
\cite{damgaard4}. The bosonization procedures are completely different
in these two papers. In the first paper the bosonic action is found
from the Jacobian of the chiral rotation, while in the second paper it
is the result of integrating out the fermion.  I will demonstrate here
that also mass term bosonization can be obtained from the
Jacobian. This will also serve as a review of the smooth bosonization
scheme. The point in performing this exercise is that it suggests that
the identification of the quantity ${\cal L}_J$ in the expression
$J=\exp(i\int d^Dx{\cal L}_J)$ for the Jacobian with the Lagrangian of
the bosonic theory is a general feature of this scheme, and may be
expected also in 4D.

The path integral with mass terms is
\begin{eqnarray}
Z[m,m^\dagger] & = & \int{\cal D}\psi{\cal D}\bar\psi
\exp i\int d^2x\bar\psi[i\dslash-mP_+-m^\dagger P_-]\psi,
\end{eqnarray}
where $P_\pm\equiv\half(1\pm\gamma_5)$ and the chiral mass $m$ is
defined from the scalar and pseudoscalar mass, $S$ and $P$, by
$m(x)\equiv S(x)+iP(x)$. The chiral change of variables
\begin{eqnarray}
\psi(x) & \to & e^{i\theta(x)\gamma_5}\psi(x), \hspace{2em}
\bar\psi(x) \;\to\; \bar\psi(x)e^{i\theta(x)\gamma_5}
\end{eqnarray}
leads to the rotated path integral
\begin{eqnarray}
\label{path}
\nonumber
Z & = & \int{\cal D}\psi{\cal D}\bar\psi
\exp i\int d^2x\bigg(\bar\psi[i\dslash-\dslash\theta\gamma_5
-me^{2i\theta}P_+-m^\dagger e^{-2i\theta}P_-]\psi \\
  & & \hspace{2em} \mbox{} +\frac{1}{2\pi}
\partial_\mu\theta\partial^\mu\theta
+\frac{1}{4\pi}\kappa_1m(e^{2i\theta}-1)
+\frac{1}{4\pi}\kappa_1m^\dagger(e^{-2i\theta}-1)+O(m^2)\bigg).
\end{eqnarray}
An appropriate regularization scheme is assumed here, one where the
current is conserved\footnote{The requirement of current conservation
alone does not completely fix the regularization scheme. The
ambiguities can be resolved by imposing Bose symmetry in triangle
diagrams instead \cite{banerjee}. The conservation of the current then
follows.}. This gives the second line in eq.\ (\ref{path}) as the
contribution from the Jacobian. $\kappa_1$ is an arbitrary massive
parameter. Note that the kinetic term and the terms proportional to
$m$ in the Jacobian is the desired result for the Lagrangian of the
bosonic theory.

We now promote the local parameter $\theta$ to a dynamical field and
thus include the path integration over this field. The path integral
does not depend on $\theta$ (at this point), so this produces an
irrelevant infinite numerical factor which is absorbed in the path
integration measure. There is now a local symmetry
\begin{eqnarray}
\psi(x) & \to & e^{i\lambda(x)\gamma_5}\psi(x), \hspace{2em}
\bar\psi(x) \;\to\; \bar\psi(x)e^{i\lambda(x)\gamma_5}, \hspace{2em}
\theta(x) \;\to\; \theta(x)-\lambda(x)
\end{eqnarray}
in the system, provided the transformation of the measure is included.

According to the smooth bosonization scheme, this local symmetry is
viewed as an ordinary gauge symmetry which can be gauge-fixed. The
gauge fixing procedure is then to choose a gauge-fixing function
$\Phi(x)$ and to insert a functional delta function
\begin{eqnarray}
\delta(\Phi) & = & \int{\cal D}\beta e^{i\int d^2x\beta\Phi}
  \;\equiv\; \int{\cal D}\beta e^{i\int d^2x{\cal L}_\mathrm{gf}}
\end{eqnarray}
and a Faddeev--Popov determinant
\begin{eqnarray}
\mathrm{Det}\left(\frac{\delta\Phi}{\delta\lambda}\right)
  & = & \int{\cal D}c{\cal D}\bar ce^{i\int d^2x
\bar c\left(\frac{\delta\Phi}{\delta\lambda}\right)c}
  \;\equiv\; \int{\cal D}c{\cal D}\bar c
e^{i\int d^2x{\cal L}_\mathrm{ghosts}}
\end{eqnarray}
into the path integral. It is necessary to include the anomalous
variation of the gauge-fixing function in order to get the correct
Faddeev--Popov determinant. In other words, it is necessary to include
the contribution from the Jacobian for the BRST variation
\begin{eqnarray}
\nonumber
\delta(-\bar c\Phi) & = & \beta\Phi
+\bar c\left(\frac{\delta\Phi}{\delta\lambda}\right)c \\
  & = & {\cal L}_\mathrm{gf}+{\cal L}_\mathrm{ghosts}
\end{eqnarray}
that is added to the Lagrangian, see the discussion in ref.\
\cite{damgaard3}.

One can now consider a general gauge-fixing function $\Phi$ which
interpolates between fermionic and bosonic variables. In this paper I
am only interested in gauges which give bosonization, thus $\Phi$
should depend on the fermion fields only. The choice I make is
\begin{eqnarray}
\label{gaugefix}
\Phi & = & \partial_\mu j_5^\mu, \hspace{2em}
j^5_\mu \;\equiv\; \bar\psi\gamma_\mu\gamma_5\psi.
\end{eqnarray}
This is the preliminary gauge-fixing chosen in ref.\ \cite{damgaard3}
(eq.\ (20) in that paper with $\Delta=1$). It is one of the
``endpoints'' of a smooth gauge; the other one is
$\Phi=\square\theta/\pi$ ($\Delta=0$ in eq.\ (20) of ref.\
\cite{damgaard3}) which returns the original fermionic
theory. Eventually, another gauge-fixing condition was adopted in
ref.\ \cite{damgaard3}, which was a formal integration of eq.\
(\ref{gaugefix}), in order to take care of certain zero
modes. However, in this paper I will assume that appropriate boundary
conditions have been imposed on the fields of the theory in order to
make all free fields vanish. Then there are only trivial zero modes
in the theory. When we set $\Phi=0$, it gives $\partial_\mu
j^\mu_5=0$, and this will then decouple the fermion from the bosonic
field $\theta$ -- except for the mass terms -- as seen from the path
integral (\ref{path}). Furthermore, in this gauge the ghosts decouple.

Thus, the problem is to decouple also the fermionic mass terms from
$\theta$. In the treatment of this problem in ref.\ \cite{damgaard2}
the massive parameter $\kappa_1$ in the Jacobian is chosen to be
zero. Instead the required terms -- linear in $m$ and $m^\dagger$ --
are generated by rewriting parts of the theory into the Schwinger
model with a ``perturbation'' and evaluating a few relevant
expectation values. However, we can solve this problem in another way
if we make use of Coleman's results \cite{coleman}. The point is that
decoupling will occur if the expression
\begin{equation}
\label{fermionic}
\int{\cal D}\beta{\cal D}\psi{\cal D}\bar\psi\exp i\int d^2x
\bar\psi[i\dslash-\dslash\beta\gamma_5
-mP_+e^{2i\theta}-m^\dagger P_-e^{-2i\theta}]\psi
\end{equation}
is unity. By using Coleman's results, we can bosonize this path
integral. We get
\begin{eqnarray}
\label{bosonic}
\nonumber
\lefteqn{\int{\cal D}\beta{\cal D}\phi\exp i\int d^2x
\bigg(\frac{1}{2\pi}\partial_\mu\phi\partial^\mu\phi
+\partial_\mu\beta\frac{1}{\pi}\partial^\mu\phi} \\
  & & \hspace{2em} \mbox{} +\frac{1}{4\pi}\kappa m
e^{2i\theta}(e^{2i\phi}-1)
+\frac{1}{4\pi}\kappa m^\dagger e^{-2i\theta}(e^{-2i\phi}-1)
+O(m^2)\bigg).
\end{eqnarray}
Here, the massive parameter $\kappa$ is arbitrary. It is clear from
consistency reasons that it must be equal to $\kappa_1$, and also that
the terms of $O(m^2)$ should be present, but this is not needed for
the argument. (The terms of $O(m^2)$ do not appear in Coleman's
expressions, probably as a consequence of his definition of the
composite operators $\sigma_\pm(x)$ in terms of point-splitting
\cite{coleman}.) We should also recall that the replacement of eq.\
(\ref{fermionic}) with eq.\ (\ref{bosonic}) holds in the sense of a
perturbative expansion in $m$ and $m^\dagger$. This is what Coleman
showed, and is a result that to our knowledge has not been improved
upon. Now it is seen that the path integration over $\beta$ gives a
delta function for $\phi$ to be a free field, hence to vanish, and so
the expression is unity.

This demonstrates that ${\cal L}_J$ is the Lagrangian of the bosonized
theory within the smooth bosonization scheme, even for massive
fermions. Unfortunately, we needed Coleman's results to prove
it. Therefore, what we just did amounts to a consistency check, and
not an independent derivation of bosonization. Similar results are not
available in 4D because it involves statements about the explicit
forms of nontrivial Green's functions. However, this has not
necessarily been a useless exercise if we can understand why the
condition for the divergence of the axial current to vanish was strong
enough to decouple the mass terms. Let us now consider this point.

First of all, let us observe that there is phase rotation invariance
in the theory, which implies a conservation equation for the current:
\begin{eqnarray}
\partial_\mu j^\mu & = & 0, \hspace{2em}
j_\mu \;\equiv\; \bar\psi\gamma_\mu\psi.
\end{eqnarray}
Note the similarity between this conservation equation and the
gauge-fixing condition. Indeed,
\begin{eqnarray}
\Psi & \equiv & \partial_\mu j^\mu
\end{eqnarray}
is itself a gauge-fixing function if we carry out the same procedure
for the phase rotations as we have just done for the chiral phase
rotations. The new ``dynamical'' field $\alpha$ then couples to the
fermion through
\begin{eqnarray}
\int{\cal D}\alpha e^{i\int d^2x\alpha\partial_\mu j^\mu}
  & = & \int{\cal D}\alpha e^{i\int d^2x\alpha\Psi}
  \;=\; \delta(\Psi),
\end{eqnarray}
and is therefore the field that implements the delta function for
$\Psi$. Strictly speaking, we have not really done any gauge-fixing,
in the sense of inserting a further delta function and an associated
Faddeev--Popov determinant, and therefore there are no ghosts.

The point is now this: Let us consider the axial current and make the
Hodge decomposition
\begin{eqnarray}
j_\mu^5 & = & \partial_\mu\xi
+\epsilon_{\mu\nu}\partial^\nu\eta.
\end{eqnarray}
(I work in a spacetime with trivial topology in this paper, so there
is no harmonic form in the decomposition.) The gauge-fixing condition
now implies
\begin{eqnarray}
\square\xi & = & 0.
\end{eqnarray}
Conversely, due to the property
\begin{eqnarray}
\gamma^\mu & = & \epsilon^{\mu\nu}\gamma_\nu\gamma_5
\end{eqnarray}
of the gamma matrices in two dimensions, current conservation implies
\begin{eqnarray}
-\epsilon_{\mu\nu}\partial^\nu \epsilon^{\mu\rho}\partial_\rho\eta
  & = & \square\eta \;=\; 0.
\end{eqnarray}
This means that effectively
\begin{eqnarray}
\square j_\mu^5=0
\end{eqnarray}
in the path integral, i.e.\ in the sense of a delta function
constraint in the path integral. In other words, the axial vector
vanishes.

We are therefore dealing with a theory where the axial current, hence
also the current, vanishes. Could this be the ``real'' reason for the
decoupling of $\theta$ from the mass terms? The following heuristic
argument suggests that it is: Since the axial current vanishes, we
have a {\em local} condition (it applies pointwise) which effectively
puts two DOFs of the fermion to zero. However, a physical Dirac
fermion in two dimensions only {\em have} two DOFs, which means that
there were not any DOFs ``left over'' for the mass terms.

This argument works in 4D as well if we can find similar local
constraints. The idea is to count the DOFs affected by these
constraints, and conclude that there is no room left for mass terms or
other bilinear combinations of the fermion field. This leads to our
strategy for smooth bosonization in 4D. The local constraints that I
will consider in 4D are to have both the vector and the axial vector
current vanish. Thus, my choice of gauge-fixing functions will be
guided by this, rather than trying to find gauges where the ghosts
decouple (which may not even exist).

\section{Smooth bosonization of an Abelian fermion in 4D}

I consider an Abelian fermion of mass $m$ coupled to an external
vector $A_\mu(x)$. The theory is described by the path integral
\begin{eqnarray}
\nonumber
Z[A] & = & \int{\cal D}\psi{\cal D}\bar\psi
e^{i\int d^4x\;{\cal L}}, \\
{\cal L} & = & \bar\psi[i\dslash-\;\fslash\! A-m]\psi.
\end{eqnarray}
I will briefly consider generalizations to non-Abelian fermions later.

First let us try the chiral phase of the fermion and see how far we
can get with bosonization in this case. Of course, our requirements
for the bosonized theory can not be fulfilled with only the chiral
phase, because topologically stable configurations can not be achieved
with just one pseudoscalar field, but we will get an idea of what is
going on. Thus, we perform the change of variables
\begin{eqnarray}
\psi & \to & e^{i\theta\gamma_5}\psi,
  \hspace{2em} \bar\psi \;\to\; \bar\psi e^{i\theta\gamma_5}
\end{eqnarray}
in the path integral, which then becomes
\begin{eqnarray}
\nonumber
Z[A] & = & J[\theta]\int{\cal D}\psi{\cal D}\bar\psi
e^{i\int d^4x\;{\cal L}'}, \\
{\cal L}' & = & \bar\psi[i\dslash-\dslash\theta\gamma_5-\;\fslash\! A
-me^{2i\theta\gamma_5}]\psi,
\end{eqnarray}
and $J[\theta]=\exp(iS_J[\theta])$ is the Jacobian of the
transformation. According to our previous considerations,
$S_J[\theta]$ is an action for the field $\theta$ and a part of the
final action for the bosonized theory. I will consider the difficult
problem of how to calculate this in the next section, but we do not
need the explicit form here.

By the smooth bosonization scheme we now path integrate over $\theta$
and insert a gauge-fixing delta function and Faddeev--Popov
determinant. An obvious choice of gauge-fixing function is
\begin{eqnarray}
\label{axial-div}
\Phi & = & \partial_\mu j^\mu_5.
\end{eqnarray}
This decouples the axial current from $\theta$, leaving only the mass
term to couple $\theta$ and the fermion.

The Faddeev--Popov determinant can now be found by adding $\beta\Phi$
to the Lagrangian and performing a gauge transformation. There is
actually a subtlety connected with this, since the new terms in the
Lagrangian is of the form
\begin{equation}
\left(\beta F_1(\theta)+\beta^2F_2(\theta)
+\beta^3F_3(\theta)\right)\delta\lambda,
\end{equation}
where $\delta\lambda$ is the parameter of the gauge
transformation. This can be found from a calculation similar to that
described in the next section. Thus it appears that the path
integration over $\beta$ no longer gives a delta function. It turns
out, however, that only $F_1$ contributes to the Faddeev--Popov
determinant:
\begin{eqnarray}
\frac{\delta\Phi}{\delta\lambda} & = & F_1(\theta).
\end{eqnarray}
This gives the BRST invariant result, and can be verified by the
alternative procedure of adding $\delta(-\bar c\Phi)$ to the
Lagrangian. $F_1$ is a derivative operator depending on $\theta$. I
will not give the explicit form here, mainly because it is tedious to
calculate, but also because the inclusion of further DOFs will modify
it, and because we will not need the explicit form. Thus the ghosts do
not decouple in this gauge. However, the meaning of the path integral
over $\beta$ as a delta function is still intact.

Let us see if the gauge-fixing condition leads to the vanishing of the
axial current $j_\mu^5$ itself. To find out we perform a Hodge
decomposition like we did in 2D. It is
\begin{eqnarray}
j_\mu^5 & = & \partial_\mu\xi
+\half\epsilon_{\mu\nu\rho\sigma}\partial^\nu\eta^{\rho\sigma}.
\end{eqnarray}
The gauge-fixing condition leads immediately to
$\square\xi=0$. $j_\mu^5$ will then be a free field -- hence vanish --
if $\eta_{\mu\nu}$ is a free field. This happens if
\begin{eqnarray}
\label{curlfree}
\half\epsilon_{\mu\nu\rho\sigma}\partial^\rho j^\sigma_5 & = & 0.
\end{eqnarray}
Using the property
\begin{eqnarray}
\label{property}
\half\{\sigma_{\mu\nu},\gamma_\rho\}
  & = & -\epsilon_{\mu\nu\rho\sigma}\gamma^\sigma\gamma_5
\end{eqnarray}
of the gamma matrices, eq.\ (\ref{curlfree}) is equivalent to
\begin{eqnarray}
\partial_\rho(\half\bar\psi
\{\half\sigma^{\mu\nu},\gamma^\rho\}\psi) & = & 0.
\end{eqnarray}
We can compare this to the equation for the conservation of the
angular momentum current:
\begin{eqnarray}
\partial_\mu j^{\mu,\alpha\beta} & \equiv & \partial_\mu(\half\bar\psi
\{\gamma^\mu,\half\sigma^{\alpha\beta}\}\psi
  + \bar\psi\gamma^\mu L^{\alpha\beta}\psi)
  \;=\; 0,
\end{eqnarray}
where $L_{\mu\nu}=x_\mu i\partial_\nu-x_\nu i\partial_\mu$ is the
orbital angular momentum operator. Thus, the spin part of this
conservation equation, together with the gauge-fixing condition, would
make the axial current a free field.

The orbital part of the conservation equation can also be written
\begin{equation}
\label{orbital}
\partial_\mu(x^\alpha\Theta^{\mu\beta}-x^\beta\Theta^{\mu\alpha}),
\end{equation}
where
\begin{eqnarray}
\Theta_{\mu\nu} & \equiv & \bar\psi\gamma_\mu i\partial_\nu\psi
\end{eqnarray}
is the energy-momentum tensor. But translation invariance implies
\begin{eqnarray}
\partial_\mu\Theta^{\mu\nu} & = & 0,
\end{eqnarray}
and the orbital part, eq.\ (\ref{orbital}), becomes
\begin{eqnarray}
\partial_\mu(x^\alpha\Theta^{\mu\beta}-x^\beta\Theta^{\mu\alpha})
  & = & \Theta^{\alpha\beta}-\Theta^{\beta\alpha}.
\end{eqnarray}
Thus, gauge-fixing and Poin\-car\'{e} invariance does not lead to the
vanishing of the axial current as a local constraint because the
energy-momentum tensor of a Dirac fermion is not symmetric.

Let us now consider the chiral Poincar\'{e} transformations. First, we
concentrate on the chiral Lorentz transformations, acting on the Dirac
fermion by
\begin{eqnarray}
\psi & \to & e^{i\frac{1}{2}\phi_{\mu\nu}
J^{\mu\nu}\gamma_5}\psi \hspace{2em}
\bar\psi \;\to\; \bar\psi e^{i\frac{1}{2}\phi_{\mu\nu}
J^{\mu\nu}\gamma_5}.
\end{eqnarray}
This is a global transformation, and a symmetry of the kinetic part of
the Lagrangian. Couplings to vectors and mass terms (and tensors)
break the symmetry explicitly. In addition there are anomalies, which
are crucial to our discussion since they are responsible for the
Jacobian.

In order to bosonize these DOFs we proceed as for the chiral phase: A
local change of variables in the path integral, with now the field
$\phi_{\mu\nu}(x)$ as parameter of the transformation, is
performed. This gives rise to a Jacobian which depends on this field,
and to which we will return in the next section. The fermionic part of
the new transformed Lagrangian is
\begin{eqnarray}
\label{transformed}
\nonumber
{\cal L}' & = & \bar\psi[i\dslash-\half\partial_\mu\phi_{\alpha\beta}
(\half\{\gamma^\mu,\half\sigma^{\alpha\beta}\}\gamma_5
+\gamma^\mu\gamma_5L^{\alpha\beta}) \\
  & & \mbox{} \hspace{2em} -\;\fslash\! A-\phi_{\mu\nu}
(A^\nu\gamma^\mu\gamma_5+x^\mu\partial^\nu\fslash\! A\gamma_5)
-m-im\phi_{\mu\nu}J^{\mu\nu}\gamma_5]\psi+\cdots.
\end{eqnarray}
I have only given the terms of lowest order in $\phi_{\mu\nu}$; the
dots refer to higher orders. It is understood that a symmetric product
of $\phi_{\mu\nu}$ and $J_{\mu\nu}$ is used, and derivative operators
are symmetrized to act both to the right and to the left, all in order
to get a real expression. From the last term in eq.\
(\ref{transformed}), the coupling term involving the mass $m$, it is
seen that $\phi_{\mu\nu}$ must have the same properties under discrete
symmetries as the bilinear combination
$\bar\psi\sigma_{\mu\nu}\gamma_5\psi$ in order not to come in conflict
with these discrete symmetries. This means that the dual tensor
$\tilde\phi_{\mu\nu}$ transforms in the same way as the
electromagnetic tensor $F_{\mu\nu}$.

The bosonic field $\phi_{\mu\nu}$ couples in a complicated way to the
fermion. But a choice for a gauge-fixing function which almost
suggests itself (by analogy to the previous coupling to $\theta$) is
\begin{eqnarray}
\Phi^{\alpha\beta} & = & \partial_\mu
(\half\bar\psi\{\gamma^\mu,\half\sigma^{\alpha\beta}\}\gamma_5\psi
+\bar\psi\gamma^\mu\gamma_5L^{\alpha\beta}\psi)
  \;\equiv\; \partial_\mu j_5^{\mu,\alpha\beta}.
\end{eqnarray}
The gauge-fixing condition leads to a manifest decoupling only between
$\psi$ and $\phi_{\mu\nu}$ in the lowest order term with
$\partial_\mu\phi_{\alpha\beta}$ in ${\cal L}'$, but the sense of the
decoupling is implicitly stronger.

The remarks about the gauge-fixing procedure for the chiral phase
apply here as well. The delta function leads to the term
\begin{eqnarray}
{\cal L}_\mathrm{gf} & = & \half\beta_{\mu\nu}\Phi^{\mu\nu}
\end{eqnarray}
in the Lagrangian, where $\beta_{\mu\nu}$ is the Lagrange
multiplier field. The Faddeev--Popov determinant gives
\begin{eqnarray}
{\cal L}_\mathrm{ghosts} & = & \half\bar c_{\mu\nu}
\left(\frac{\delta\Phi^{\mu\nu}}{\delta\lambda^{\rho\sigma}}\right)
c^{\rho\sigma},
\end{eqnarray}
where $\delta\Phi^{\mu\nu}/\delta\lambda^{\rho\sigma}$ is a
complicated derivative operator depending on $\phi_{\mu\nu}$.

Let us investigate the spin part of the gauge-fixing condition. It is
\begin{eqnarray}
\partial_\mu(\half\bar\psi\{\gamma^\mu,
\half\sigma^{\alpha\beta}\}\gamma_5\psi) & = & 0,
\end{eqnarray}
which is equivalent to
\begin{eqnarray}
\label{spin}
\half\epsilon^{\mu\alpha\beta\nu}\partial_\mu j_\nu
  & = & 0
\end{eqnarray}
due to the property (\ref{property}).  The meaning of this equation is
revealed if we Hodge decompose the vector $j_\mu$:
\begin{eqnarray}
j_\mu & = & \partial_\mu\xi+\half\epsilon_{\mu\nu\rho\sigma}
\partial^\nu\eta^{\rho\sigma}.
\end{eqnarray}
From this we can deduce that eq.\ (\ref{spin}) implies
$\square\eta_{\mu\nu}=0$. Moreover, from phase rotation invariance, we
have the current conservation equation
\begin{eqnarray}
\partial_\mu j^\mu & = & 0,
\end{eqnarray}
from which also $\square\xi=0$ follows. Thus, modulo the orbital part
of the gauge-fixing condition, we have effectively a delta function in
the path integral which ensures that the current $j_\mu$ is a free
field.

Then what about the orbital part? This reads
\begin{eqnarray}
\label{orbital-div}
\partial_\mu(\bar\psi\gamma^\mu\gamma_5
L^{\alpha\beta}\psi)
  & = & \partial_\mu(x^\alpha\Theta_5^{\mu\beta}
-x^\beta\Theta_5^{\mu\alpha}),
\end{eqnarray}
where I have introduced the ``axial energy-momentum tensor''
\begin{eqnarray}
\Theta_5^{\mu\nu}
 & \equiv & \bar\psi\gamma^\mu\gamma_5i\partial^\nu\psi.
\end{eqnarray}

At this point we introduce further DOFs in the theory, corresponding
to chiral translations. Thus we make the change of variables
\begin{eqnarray}
\psi & \to & e^{ib_\mu P^\mu\gamma_5}\psi, \hspace{2em}
\bar\psi \;\to\; \bar\psi e^{ib_\mu P^\mu\gamma_5}
\end{eqnarray}
in the path integral. This leads to a Jacobian and to the new
Lagrangian
\begin{eqnarray}
\nonumber
{\cal L}' & = & \bar\psi[i\dslash
-\partial_\mu b_\nu\gamma^\mu\gamma_5\partial^\nu \\
  & & \mbox{} \hspace{2em} -\;\fslash\! A
-b_\mu\partial^\mu\;\fslash\! A\gamma_5
-m-2imb_\mu P^\mu\gamma_5]\psi+\cdots.
\end{eqnarray}
The field $b_\mu$ is an axial vector.

To fix the new gauge symmetry we choose the condition
\begin{eqnarray}
\Phi^\nu & = & \partial_\mu
(\bar\psi\gamma^\mu\gamma_5i\partial^\nu\psi)
  \;=\; \partial_\mu\Theta_5^{\mu\nu} \;=\; 0.
\end{eqnarray}
The new terms in the Lagrangian from the gauge-fixing procedure are
\begin{eqnarray}
{\cal L}_\mathrm{gf} & = & \beta_\mu\Phi^\mu
\end{eqnarray}
and
\begin{eqnarray}
{\cal L}_\mathrm{ghosts} & = & \bar c_\mu
\left(\frac{\delta\Phi^\mu}{\delta\lambda^\nu}\right)c^\nu.
\end{eqnarray}
In the presence of the delta function for the gauge-fixing condition,
the orbital part of the chiral Lorentz gauge-fixing function, eq.\
(\ref{orbital-div}), effectively becomes
\begin{eqnarray}
\Theta_5^{\mu\nu}-\Theta_5^{\nu\mu}.
\end{eqnarray}
Thus, an obstacle for having effectively a vanishing current $j_\mu$
is the antisymmetric part of $\Theta_5^{\mu\nu}$.

We have now included all the chiral DOFs advertized in the
Introduction in the path integral. Putting all together, we come to
the following result:
\begin{eqnarray}
Z & = & \int{\cal D}\psi{\cal D}\bar\psi
{\cal D}[a]{\cal D}[b]{\cal D}[c]{\cal D}[\bar c]
e^{i\int d^4x{\cal L}_\mathrm{eff}}
\end{eqnarray}
where
\begin{eqnarray}
\nonumber
{\cal D}[a] & \equiv & {\cal D}\theta\,
{\cal D}\phi_{\mu\nu}\,{\cal D}b_\mu, \\
\nonumber
{\cal D}[b] & \equiv & {\cal D}\beta\,
{\cal D}\beta_{\mu\nu}\,{\cal D}\beta_\mu, \\
{\cal D}[c]{\cal D}[\bar c] & \equiv & {\cal D}c\,{\cal D}\bar c\,
{\cal D}c_{\mu\nu}\,{\cal D}\bar c_{\mu\nu}\,
{\cal D}c_\mu\,{\cal D}\bar c_\mu,
\end{eqnarray}
and the effective Lagrangian is
\begin{eqnarray}
{\cal L}_\mathrm{eff} & = & {\cal L}'+{\cal L}_\mathrm{gf}
+{\cal L}_\mathrm{ghosts}+{\cal L}_J,
\end{eqnarray}
where
\begin{eqnarray}
\label{eff-lagrangian}
\nonumber
{\cal L}' & = & \bar\psi e^{iB\gamma_5}[i\dslash-\;\fslash\! A-m]
e^{iB\gamma_5}\psi, \hspace{2em}
B \;\equiv\; \theta+\half\phi_{\mu\nu}J^{\mu\nu}+b_\mu P^\mu, \\
\nonumber
{\cal L}_\mathrm{gf} & = & \beta\Phi
+\half\beta_{\mu\nu}\Phi^{\mu\nu}+\beta_\mu\Phi^\mu, \\
\nonumber
{\cal L}_\mathrm{ghosts} & = & \bar c\left(\frac{\delta\Phi}
{\delta\lambda}\right)c
+\half\bar c_{\mu\nu}\left(\frac{\delta\Phi^{\mu\nu}}
{\delta\lambda^{\rho\sigma}}\right)c^{\rho\sigma}
+\bar c_\mu\left(\frac{\delta\Phi^\mu}
{\delta\lambda^\nu}\right)c^\nu, \\
\int d^4x{\cal L}_J & = & -i\ln J[B].
\end{eqnarray}
In addition, the three conservation equations (current, angular
momentum and energy-momen\-tum) implicitly hold. They can be included
explicitly by introducing the Lagrange multiplier $\alpha$,
$\alpha_{\mu\nu}$ and $\alpha_\mu$, and adding
\begin{equation}
\alpha\partial_\mu j^\mu
+\half\alpha_{\mu\nu}\partial_\rho j^{\rho,\mu\nu}
+\alpha_\mu\partial_\nu\Theta^{\nu\mu}
\end{equation}
to the effective Lagrangian.

It is now clear that exact bosonization fails, at least with only the
chiral DOFs that we have considered. The currents do not vanish, and
the ghosts do not decouple. The meaning of this result, and the
question of how far we are from exact bosonization, will be discussed
in sec.\ 5.

\section{The bosonic action}

I now turn to the problem of calculating the quantity ${\cal L}_J$
from the Jacobian $J=\exp(i\int d^4x{\cal L}_J)$ of the chiral
transformations. This is the Lagrangian for the would-be bosonized
theory.

Technically, the full change of variables
\begin{eqnarray}
\label{full-change}
\psi & \to & e^{iB\gamma_5}\psi, \hspace{2em}
\bar\psi \;\to\; \bar\psi e^{iB\gamma_5},
\end{eqnarray}
with $B$ as in eq.\ (\ref{eff-lagrangian}), is difficult to
implement. We may anticipate that the action $S_J$ is a complicated
nonlinear functional of the bosonic fields, containing infinitely many
derivatives and angular momentum operators.  Approximations are
therefore necessary. Observe that the latter features come from the
derivatives in the infinitesimal generators, that is, their spacetime
parts. The first approximation I will adopt for my calculations is to
ignore these parts of the generators, that is, I will ignore the
orbital part of $J_{\mu\nu}$ in the term with $\phi_{\mu\nu}$, and the
field $b_\mu$ altogether. Intuitively this is a low energy
approximation.

Still further simplifications are necessary. I will consider the
action for $\theta$ and $\phi_{\mu\nu}$ separately to begin with;
interactions between the fields will be investigated later.

The Jacobian $J[\theta]=\exp(iS_J[\theta])$ can be calculated {\em
exactly} provided the regularization scheme is carefully chosen. There
are two things we must pay special attention to. The first one is that
phase and Poincar\'{e} invariance must be respected. We can achieve
this by using the proper time regularization scheme described in ref.\
\cite{thomassen}. This is not an unimportant point, because an axial
vector will appear in the Dirac operator in intermediate calculations,
and this may destroy Poincar\'{e} invariance for certain schemes, see
ref.\ \cite{thomassen}. But other schemes should also be possible. The
second thing is that proper time regularization by itself is
quadratically divergent. For our calculations we have removed the
quadratic divergences by extending the regularized proper time
integral into a Pauli--Villars sum \cite{pauli}. The procedure is
described in ref.\ \cite{hu}, where it is used for the calculation of
chiral anomalies in the path integral.

The calculation proceeds by integrating up a sequence of infinitesimal
chiral rotations, and the result is
\begin{eqnarray}
\label{theta}
\nonumber
{\cal L}_\theta & = & \frac{1}{24\pi^2}\theta\,\square^2\,\theta
+\frac{1}{12\pi^2}(\partial_\mu\theta\partial^\mu\theta)^2 \\
\nonumber
  & & \mbox{} -\frac{m^2}{12\pi^2}
\partial_\mu\theta\partial^\mu\theta
-\frac{m^2}{6\pi^2}\left(e^{4i\theta}+e^{-4i\theta}\right)
\partial_\mu\theta\partial^\mu\theta \\
\nonumber
  & & \mbox{} -\frac{m^4}{24\pi^2}\left(e^{4i\theta}
+e^{-4i\theta}-2\right)
-\frac{m^4}{192\pi^2}\left(e^{8i\theta}+e^{-8i\theta}-2\right) \\
  & & \mbox{} +\frac{1}{8\pi^2}\theta F\tilde F,
\end{eqnarray}
with $S_J=\int d^4x{\cal L}_\theta$ and $F_{\mu\nu}=\partial_\mu
A_\nu-\partial_\nu A_\mu$. It is a kind of generalized sine--Gordon
Lagrangian; the last term is the familiar Adler--Bell--Jackiw (ABJ)
anomaly. I repeat that this is an exact result. There is no hidden
dependence on the regularization scheme in terms of carefully chosen
parameters, and the cutoff has been taken to infinity. Furthermore,
there are no higher powers of derivatives than four.

I will return to the physics of ${\cal L}_\theta$ in the next section,
but let us here expand it in powers of $\theta$ and keep only the
Gaussian part -- a weak field approximation. We also ignore the
coupling to the vector field. We get
\begin{eqnarray}
{\cal L}_\theta
  & = & \frac{1}{24\pi^2}\theta\,\square^2\,\theta
+\frac{5m^2}{12\pi^2}\theta\,\square\,\theta
+\frac{m^4}{\pi^2}\theta^2+\cdots.
\end{eqnarray}
It is more convenient to rewrite this as
\begin{eqnarray}
\label{gaussian}
\nonumber
{\cal L}_\theta & = & \frac{1}{24\pi^2}\theta\left(\square^2
+10m^2\square+24m^4\right)\theta+\cdots \\
  & = & \frac{1}{24\pi^2}\theta(\square+6m^2)
(\square+4m^2)\theta+\cdots.
\end{eqnarray}
Thus $\theta$ has a negative metric, assuming the sign of the metric
is defined from the Gaussian part of the action. Let us now try to
insert an $i\epsilon$ according to the usual prescription $m\to
m-i\epsilon$. This appears to lead to a negative imaginary part for
the Lagrangian, hence jeopardizing the convergence of the path
integral. However, we should really consider the full Lagrangian
(\ref{theta}), which contains several oscillating terms. These are
harmless with respect to divergence, and the only potentially
dangerous term is the third one,
\begin{eqnarray}
\mbox{} -\frac{m^2-i\epsilon}{12\pi^2}\partial_\mu\theta
\partial^\mu\theta.
\end{eqnarray}
But if we take a plane wave $\theta(x)=e^{ikx}$ for $\theta$,
then this term becomes
\begin{eqnarray}
\mbox{} -\frac{m^2-i\epsilon}{12\pi^2}\,k^2,
\end{eqnarray}
which leads to convergence for $k_\mu$ time-like, as is expected
on-shell. Thus the theory is well defined despite the negative metric
of $\theta$, although this may severely restrict the physical r\^{o}le
of this field.

We can in principle perform a similar analysis for the field
$\phi_{\mu\nu}$ when the orbital part of $J_{\mu\nu}$ is ignored, but
in practice our level of technology limits what we can do. A full
nonlinear Lagrangian like (\ref{theta}) would be very hard to
calculate. It is, however, possible to find the Gaussian part. A
procedure for this is first to perform one infinitesimal chiral
Lorentz transformation and calculate the Jacobian, then perform a
second transformation and calculate the Jacobian for that with the new
Dirac operator, this time keeping two orders of $\phi_{\mu\nu}$. The
correct, ``total'' Jacobian is the sum of these, but with the terms
proportional to $\phi^2$ divided by two. This procedure can be proved
by considering the total Jacobian as an integral over a sequence of
infinitesimal contributions. The Gaussian approximation for the
$\phi_{\mu\nu}$ Lagrangian is
\begin{eqnarray}
\label{phi}
{\cal L}_\phi & = & \frac{1}{192\pi^2}\phi_{\mu\nu}(\square+6m^2)
(\square+4m^2)\phi^{\mu\nu}+\frac{1}{48\pi^2}\phi_{\mu\nu}
(\square+6m^2)\tilde F^{\mu\nu}.
\end{eqnarray}
The last term is the spin part of the anomaly from ref.\
\cite{thomassen}.

Note the similarity between the two Gaussian Lagrangians for $\theta$
and $\phi_{\mu\nu}$, apart from the coupling to the vector field
$A_\mu$. In fact, we can get the latter Lagrangian by making the
replacement $\theta\to\half\phi_{\mu\nu}\half\sigma^{\mu\nu}$ and
taking the Dirac trace: $\frac{1}{4}\tr(\cdots)$. This suggests that
writing all the terms in the Lagrangian (\ref{theta}) in terms of
exponentials (i.e.\ also the kinetic terms), making the replacement
\begin{eqnarray}
\label{replacement}
\theta & \to & \theta+\half\phi_{\mu\nu}J^{\mu\nu}+b_\mu P^\mu,
\end{eqnarray}
and taking the Dirac trace divided by four leads to a quantity that is
a part of the full bosonic Lagrangian. Additional terms are expected
to appear in a ``real'' calculation from the derivative operators
$L_{\mu\nu}$ and $P_\mu$, and various terms containing the vector
field $A_\mu$. Furthermore, there may be a term similar to the
Wess--Zumino term in the chiral Lagrangian of the strong interactions,
because the $\half\sigma_{\mu\nu}$ generate a non-Abelian group, like
the Gell-Mann matrices $\half\lambda^a$ of $SU(3)$ in the usual case.

Finally, we can calculate the interaction terms between $\theta$ and
$\phi_{\mu\nu}$. There are several terms which couples one $\theta$
with one $\phi_{\mu\nu}$. These can be computed in the same way as the
Gaussian Lagrangian for $\phi_{\mu\nu}$ above, and are therefore the
only ones I will consider. It has the complicated form
\begin{eqnarray}
\nonumber
{\cal L}_{\theta\phi} & = & \frac{1}{24\pi^2}
\bigg(\frac{1}{\square}\partial^\rho F_{\rho\nu}\phi^{\nu\mu}
(\square+6m^2)\partial_\mu\theta \\
\nonumber
  & & \mbox{} \hspace{3em} +3F^{\rho\nu}\theta
\partial^\mu\partial_\rho\phi_{\nu\mu}
+\mbox{$\frac{3}{2}$}F_{\mu\nu}\theta\,\square\,\phi^{\mu\nu}
-2F^{\rho\nu}\partial^\sigma\theta\partial_\rho\phi_{\nu\sigma}
+2F^{\rho\nu}\partial_\nu\partial^\mu\phi_{\mu\rho} \\
  & & \mbox{} \hspace{3em} +F^{\rho\nu}
\partial^\sigma\partial_\rho\theta\phi_{\nu\sigma}
+\mbox{$\frac{3}{2}$}F_{\mu\nu}\,\square\,\theta\phi^{\mu\nu}
+12m^2F_{\mu\nu}\theta\phi^{\mu\nu}\bigg)
\end{eqnarray}
To obtain this result I have made the replacement
\begin{eqnarray}
A_\mu & \to & \Pi_{\mu\nu}A^\nu \;\equiv\; \left(g_{\mu\nu}
-\frac{\partial_\mu\partial_\nu}{\square}\right)A^\nu
  \;=\; \frac{1}{\square}\partial^\nu F_{\nu\mu},
\end{eqnarray}
where the projection operator $\Pi_{\mu\nu}$ removes the gradient part
of $A_\mu$ and thus renders the expression gauge invariant. This can
be done without loss of generality, since the gradient part of $A_\mu$
does not couple to the fermion. The presence of an $F_{\mu\nu}$ could
be anticipated since a nonvanishing expression could not be obtained
from one $\theta$, one $\phi_{\mu\nu}$, and derivatives alone.

This completes my calculation of the bosonic action. Two more terms
are actually known: A coupling term between one $\phi_{\mu\nu}$ and
three $A_\mu$'s, and a coupling term between one $b_\mu$ and three
$A_\mu$'s. These are part of the chiral Poincar\'{e} anomalies, and
are found in ref.\ \cite{thomassen}. They are not given here, because
they contain an $L_{\mu\nu}$ and a $P_\mu$, respectively.

It should also be possible to calculate the Faddeev--Popov
determinants in ${\cal L}_\mathrm{ghosts}$ to the same precision. This
would require only a Jacobian for an infinitesimal transformation, but
the change (\ref{full-change}) in the Lagrangian, which clearly
complicates the calculation.

\section{Discussion}

\subsection{Summary}

Let us first recall how our initial requirements are met:

(i) the generalized chiral phases are the antisymmetrical tensor field
$\phi_{\mu\nu}$ from the chiral Lorentz transformations, and an axial
vector $b_\mu$ from the chiral translations; both of these can be used
for making configurations of nontrivial topology,

(ii) the chiral phase rotation and chiral Poincar\'{e} transformations
give rise to a Jacobian when a change of variables are made in the
path integral, and

(iii) the quantity ${\cal L}_J$ is at least a part of the Lagrangian
for the bosonic theory -- if bosonization had been exact, it would be
the complete Lagrangian; it is nonlinear, hence possibly have soliton
solutions, and contains fourth order derivative terms, which can
stabilize these solitons.

The question of whether or not topologically stable solitons can be
formed in the bosonic theory is interesting by itself, and deserves
further investigation. I favor the possibility that the tensor
$\phi_{\mu\nu}$ is responsible for these configurations, rather than
the axial vector $b_\mu$. The reason is that $b_\mu$ is the parameter
of a pure derivative operator, $P_\mu\gamma_5$, while $\phi_{\mu\nu}$
is the parameter of $J_{\mu\nu}\gamma_5$ which at least has an
internal part, $\half\sigma_{\mu\nu}\gamma_5$. Hence, if part of the
Lagrangian for $\phi_{\mu\nu}$ is indeed given by the replacement
(\ref{replacement}) in eq.\ (\ref{theta}), then that part bears a
resemblance to the Skyrme model where the matrices $\sigma_{\mu\nu}$
play the r\^{o}le of the Pauli or Gell-Mann matrices.

Guided by the observation that the axial vector current (and vector
current) vanishes in a certain gauge of the smooth bosonization
scheme, I tried in this paper to find the gauges which would implement
the vanishing of the vector and axial vector currents in 4D. However,
this program was not completely successful due to the fact that the
energy-momentum tensor for a Dirac fermion, and its axial counterpart,
is not symmetrical. Furthermore, in the chosen gauge the ghosts did
not decouple. Nevertheless, this is an exact rewriting of the theory,
exhibiting some DOFs of the fermion in an unusual way.

\subsection{Problems with the approach}

If what we desire is exact bosonization, the most serious problem is
that the ghosts do not decouple from the bosonic fields. Even if we
did get rid of the fermion, the ghosts would be there as an obstacle
for this. Conversely, if we could find a ghost-free gauge -- and it is
far from clear that this exists -- then the {\em fermion} would
probably not be decoupled. Thus exact bosonization must fail, at least
for the model and DOFs I have considered in this paper.

Another problem is the lack of symmetry of the energy-momentum
tensors. However, in applications the energy-momentum tensor is
frequently replaced by a symmetrical improved tensor. This suggests
that the antisymmetrical part is somehow unimportant. If we were
allowed to replace the two energy-momentum tensors with their
symmetrized versions, this would mean that the current and axial
current would effectively be free fields, hence vanish, and the
fermion would be completely eliminated from the theory, according to
our previous arguments. In this case the resulting theory would be a
theory of ``only'' bosons and ghosts.

Of course, also a problem is the lack of a proof that vanishing
currents imply a vanishing fermion. Indeed, if such a proof could be
found, it is likely to be known first in 2D. However, no such proof is
known in 2D.

These problems may imply the possibility that we have really been
going the ``wrong direction''. Perhaps it is some complicated
nonlinear bosonic theory which is the fundamental thing, and that when
``fermionized'' gives a theory of a fermion coupled to bosonic
fields. This is the situation with the Skyrme model \cite{skyrme}. If
this possibility is correct, then it would explain why an arbitrary
fermion could not be bosonized, and it would imply that certain
theories of fermions coupled to bosons {\em could}.

There is also a potential problem that may occur at finite
temperature. In the literature \cite{luther} there exists an
argument\footnote{I thank the referee for pointing out this argument.}
that each fermion DOF has the same energy as 7/8 bosonic DOFs. This
seems to restrict the construction of bosonic theories from fermionic
ones. However, it is important to take into account that the
``7/8-rule'' only applies if the system is in a state of an
approximately free gas of particles, both with respect to the
fermionic description and the bosonic description. This is not the
case for the bosonization procedure in this paper, because the
bosonized theory is nonlinear, hence cannot describe a free gas. The
same is of course true for other bosonization schemes that produces
nonlinear bosonic theories.

\subsection{Non-Abelian fermions}

I now briefly consider the possibility of generalizing the results to
a non-Abelian Dirac fermion. (See also refs.\
\cite{damgaard2,thomassen}.) The fermion will then transform in some
representation of the non-Abelian group. If we for simplicity choose
$U(N)$ and the fundamental representation, we have the ``color''
transformations
\begin{eqnarray}
\psi & \to & e^{i\omega^at^a}\psi, \hspace{2em}
\bar\psi \;\to\; \bar\psi e^{-i\omega^at^a},
\end{eqnarray}
where $a=1,\ldots,N$ and $t^a$ are the generators. This is of course a
generalization of the phase rotation of the Abelian phase, and as such
should be considered together with the Poincar\'{e} transformations
and their chiral counterparts. We must therefore admit the possibility
that also Poincar\'{e} transformations, chiral phase rotations and
chiral Poincar\'{e} transformations can be generalized to color
space. That is, each color of fermion can be transformed
separately. Thus we have the generators
\begin{equation}
t^a, \hspace{1em} t^aJ_{\mu\nu}, \hspace{1em} t^aP_\mu, \hspace{1em}
t^a\gamma_5, \hspace{1em} t^aJ_{\mu\nu}\gamma_5
\hspace{1em} \mbox{and} \hspace{1em} t^aP_\mu\gamma_5,
\end{equation}
respectively, for these transformations. Note that they do not
generate a group, since their algebra does not close. They are also
not in general symmetries of the fermion Lagrangian. However, they
have the property that they are naive symmetries of the fermionic
measure while giving rise to a Jacobian in the regularized theory. It
is for this reason that they are important for smooth bosonization.

We can now follow the same procedure as for the Abelian case,
introducing the new DOFs in the path integral, gauge-fixing etc. The
bosonic action will again be $S_J$ from the Jacobian with bosonic
fields $\theta^a$, $\phi_{\mu\nu}^a$ and $b_\mu^a$. The action will
then have further complications from a non-Abelian structure. The
presence of colored matrices in the Dirac operator will break the
color symmetries and may change the conservation equations needed for
our previous arguments. However, the new conservation equations will
probably be equally effective, and a generalization of the results of
sec.\ 3 should be straightforward.

\subsection{The physics of ${\cal L}_J$: A model theory}

Even if bosonization is not exact, it may be possible to use the
rewritten theory as a starting point for an approximate description of
certain processes, perhaps at low energies. This may be relevant in
particular when the ``effective DOFs'' in these processes are
pseudoscalars, tensors or axial vectors. But I will now consider
another approximation, motivated instead by {\em simplicity}. Namely,
I will consider the theory described by the Lagrangian ${\cal L}_J$ of
the bosonic theory in its own right. I can think of no reason why this
theory cannot be a perfectly healthy quantum field theory. In a sense
it is like a caricature of the true fermionic system. A heuristic
argument for this is given in terms of a physical interpretation of
the Jacobian $J$ below.

First, let us assume that the vector field $A_\mu$ is absent. In the
weak field limit we can find the equation of motion for $\theta$ from
eq.\ (\ref{gaussian}):
\begin{eqnarray}
(\square+6m^2)(\square+4m^2)\theta & = & 0,
\end{eqnarray}
a ``double'' Klein--Gordon equation. This has plane wave solutions:
\begin{eqnarray}
\theta(x) & = & e^{ikx}, \hspace{2em} k^2 \;=\; m_1^2\;\equiv\; 4m^2
  \hspace{1em} \mbox{or} \hspace{1em} k^2 \;=\; m_2^2\;\equiv\; 6m^2.
\end{eqnarray}
Thus there are apparently two mass shells for the field, corresponding
to two ``branches'' of propagation. The same is true for weak
$\phi_{\mu\nu}$'s, as can be seen from eq.\ (\ref{phi}). Of course,
these modes may turn out to be irrelevant when higher orders of the
fields and interactions between them are taken into account.

If we now include the vector $A_\mu$, and furthermore, assume that
this field is a dynamical field in a larger theory, then the situation
becomes more interesting. The $\theta$ may then decay into two
$A_\mu$'s through the ABJ anomaly, and the $\phi_{\mu\nu}$ mixes with
$A_\mu$ ($\phi_{\mu\nu}$ and $b_\mu$ decays also into {\em three}
$A_\mu$'s through the anomaly terms we mentioned in the previous
section). It also induces further couplings between $\theta$ and
$\phi_{\mu\nu}$ (and $b_\mu$) by $A_\mu$-exchange.

Finally, I will discuss the physical interpretation for the use of the
Jacobian of chiral transformations in the bosonic theory. We can
define the effective action $W$ for the fermionic theory by
\begin{eqnarray}
e^{iW} & \equiv & Z \;=\; \int{\cal D}\psi{\cal D}\bar\psi
e^{i\int d^4x\bar\psi D\psi},
\end{eqnarray}
where $D$ is the Dirac operator. A change of variables will give
\begin{eqnarray}
Z & = & J[B]\int{\cal D}\psi{\cal D}\bar\psi
e^{i\int d^4x\bar\psi(e^{iB\gamma_5}De^{iB\gamma_5})\psi}
  \;=\; J[B]e^{iW[B]},
\end{eqnarray}
with $B$ defined in eq.\ (\ref{eff-lagrangian}) and $W[B]$ the rotated
effective fermion action, so that the Jacobian can be written
\begin{eqnarray}
J[B] & = & \frac{\int{\cal D}\psi{\cal D}\bar\psi
e^{i\int d^4x\bar\psi D\psi}}{\int{\cal D}\psi{\cal D}\bar\psi
e^{i\int d^4x\bar\psi e^{iB\gamma_5}De^{iB\gamma_5}\psi}}
  \;=\; e^{i(W-W[B])}.
\end{eqnarray}
The quantity $S_J[B]$ in $J=e^{iS_J[B]}$ is therefore the difference
between the effective action of an {\em unrotated} fermion field and a
{\em rotated} fermion field. In a sense, $S_J[B]$ measures the
response of the fermion to external ``forces'' -- it is the amount of
action we must inject into the system to maintain the rotated
configurations $e^{iB\gamma_5}\psi$, $\bar\psi e^{iB\gamma_5}$
compared to the unrotated ones $\psi$, $\bar\psi$. An analogy which
comes to mind is the physics of an elastically deformable solid, where
the fermion field is like the solid, and the chiral transformations
are like compression, shear and tension deformations. The Lagrangian
${\cal L}_J$ thus describes the theory of such ``deformations'' of the
fermion field.

Ideas reminiscent of these have been used in ref.\ \cite{diakonov} to
justify a derivation of the chiral Lagrangian \cite{gasser} from QCD.
In these papers, however, mainly the chiral flavor phase DOFs of the
quarks were considered. It would be interesting to try to bosonize
{\em all} DOFs of the quark field, including both color and chiral
Poincar\'{e} phases. (For a related investigation see ref.\
\cite{damgaard2}.)

\noindent
\paragraph{Acknowledgments} I thank P.H.\ Damgaard for invaluable
discussions and comments, and M.\ Faber, A.N.\ Ivanov, S.H.\ Hansen,
and O.\ Borisenko for discussions.

\end{document}